\documentclass[aps,prd,notitlepage,nofootinbib,superscriptaddress,twocolumn,groupedaddress,showpacs]{revtex4}

\usepackage{graphicx,color}
\usepackage{amstext,amsmath,amssymb,amsfonts,bbm}
\usepackage{amsthm}
\usepackage{graphicx}
\usepackage{dcolumn}
\usepackage{bm}

\newcommand{\Ref}[1]{(\ref{#1})}

\newcommand{\Z}{\mathbb{Z}}
\newcommand{\R}{\mathbb{R}}
\newcommand{\C}{\mathbb{C}}

\newcommand{\half}{\frac{1}{2}}

\newcommand{\Slc}{\mathrm{SL}(2,\mathbb{C})}

\newcommand{\Su}{\mathrm{SU}(2)}

\def\be{\begin{eqnarray}}
\def\ee{\end{eqnarray}}

\newcommand{\cm}{\mathcal M}

\newcommand{\cp}{\mathcal P}

\newcommand{\cs}{\mathcal S}

\newcommand{\fp}{\mathfrak{p}}

\renewcommand{\a}{\alpha}
\renewcommand{\b}{\beta}
\newcommand{\g}{\gamma}
\newcommand{\G}{\Gamma}

\newcommand{\sig}{\sigma}

\renewcommand{\L }{\Lambda}

\renewcommand{\O}{\Omega}
\renewcommand{\t}{\tau}

\newcommand{\rmd}{\mathrm d}

\newcommand{\lt}{\left}
\newcommand{\rt}{\right}

\newcommand{\lag}{\left\langle}
\newcommand{\rag}{\right\rangle}

\newcommand{\tr}{\mathrm{tr}}

\begin{document}

\sloppy

\title{SU(2) Flat Connection on Riemann Surface and 3D Twisted Geometry with Cosmological Constant}

\author{Muxin Han}
\affiliation{Department of Physics, Florida Atlantic University, 777 Glades Road, Boca Raton, FL 33431, USA}
\affiliation{Institut f\"ur Quantengravitation, Universit\"at Erlangen-N\"urnberg, Staudtstr. 7/B2, 91058 Erlangen, Germany}

\author{Zichang Huang}
\affiliation{Department of Physics, Florida Atlantic University, 777 Glades Road, Boca Raton, FL 33431, USA}


\begin{abstract}

Twisted geometries are understood to be the discrete classical limit of Loop Quantum Gravity. In this paper, SU(2) flat connections on (decorated) 2D Riemann surface are shown to be equivalent to the generalized twisted geometries in 3D space with cosmological constant. Various flat connection quantities on Riemann surface are mapped to the geometrical quantities in discrete 3D space. We propose that the moduli space of SU(2) flat connections on Riemann surface generalizes the phase space of twisted geometry or Loop Quantum Gravity to include the cosmological constant.

\end{abstract}


\maketitle

\section{\label{sec:level1}Introduction}

There has been significant development recently on implementing cosmological constant in Loop Quantum Gravity (LQG) \cite{Haggard:2014xoa,2CCS,4dqg,Haggard:2015yda,Han:2011aa,Han:2010pz,Fairbairn:2010cp}. A new covariant formulation of LQG has been developed, and presented a nice relation between LQG in 4D and Chern-Simons theory on 3-manifolds. As one of the key results in the new formulation, the new LQG spinfoam amplitude constructed by using Chern-Simons theory reproduces correctly the 4D (discrete) Einstein-Hilbert action with cosmological constant in the semiclassical limit. 

It is interesting to understand the relation between the new covariant formulation and canonical LQG (see e.g.\cite{Han:2005km,book,review1} for reviews on canonical LQG). This paper makes the first step toward this direction. In this work we formulate a new phase space of LQG, which suitably generalizes the LQG phase space to the situation with cosmological constant. Given that the covariant LQG in 4D has been related to Chern-Simons theory on 3-manifold, the new LQG phase space defined on the 3D spatial slice relates to the phase space of Chern-Simons theory, which is the moduli space of flat connections on 2D Riemann surface. 

A useful geometrical parametrization of the phase space underlining LQG\footnote{The phase space used for LQG on the continuum is discussed in \cite{Ashtekar:1998ak}. The truncated phase space for LQG on a fixed graph is proposed in e.g.\cite{Thiemann:2000bw,Sahlmann:2001nv}.} is known as \emph{twisted geometry} \cite{LaurentFreidel2010,Langvik:2016hxn,CarloRovelli2010,Haggard:2012pm,Freidel:2010bw,Livine:2011vk,Speziale:2012nu}. The twisted geometry variables are coordinates of LQG phase space, which are interpreted as (discrete) geometrical quantities of 3D space. The twisted geometry is a classical geometry defined on a 3D triangulation (or cellular decomposition), or equivalently on a graph dual to the triangulation. As the main difference from 3D Regge geometry, the twisted geometry only has the area-matching but doesn't force the shape-matching in gluing a pair of flat tetrahedra (or polyhedra). In addition, there is a twist angle $\xi$ appearing in the gluing, which relates to the extrinsic curvature of the space (the hyper-dihedral angle between the pair of tetrahedra). 

A natural question to ask is whether the new phase space of LQG with cosmological constant also admits a geometrical parametrization, which relates to the twisted geometry. The answer is surely positive. Indeed the new phase space relates to a generalized twisted geometry in presence of cosmological constant. The generalized twisted geometry is made by tetrahedra with constant curvature\footnote{Tetrahedra with constant curvature are discussed in e.g.\cite{Bahr:2009qd} in the non-twisted-geometry context.}. As a key point, since the new LQG phase space relates to flat connections on Riemann surface, what we derive in this work is understood as the relation between flat connection on 2D surface and twisted geometry in 3D physical space. 

This work generalizes the result in \cite{Hal.M.Haggard2015}, where the authors find the relation between SU(2) flat connections on 4-holed sphere and constant curvature tetrahedron geometries \footnote{A proposal using different group is developed in \cite{Charles:2016xzi,Charles:2015lva} for the hyperbolic tetrahedron geometry.}.

The main results of this paper are summarized as follows: The truncated LQG phase space or twisted geometry is defined on a graph dual to 3D triangulation. We define in Section \ref{GRS} an 1-to-1 correspondence between a graph and a Riemann surface $\cs$ with certain decoration. The new LQG phase space is defined to be the moduli space of SU(2) flat connection on $\cs$, denoted by $\cm_{flat}(\cs,\Su)$ \footnote{$\cm_{flat}(\cs,\Su)$ is the space of SU(2) connections $A$ on $\cs$ with vanishing curvature $F_A=0$, quotient by SU(2) gauge transformations. $\cm_{flat}(\cs,\Su)$ is in general finite dimensional. Flat connections in $\cm_{flat}(\cs,\Su)$ can be parametrized by holonomies along loops in $\cs$. Fixing the base point of a loop, the continuous deformation of the loop doesn't affect the loop holonomy, since the connection is flat. So the loop holonomy only depends on the homotopy class of loops. Therefore $\cm_{flat}(\cs,\Su)$ is isomorphic to the space of loop holonomies over the fundamental group $\pi_1(\cs)$, quotient by gauge transformations. Namely $\cm_{flat}(\cs,\Su)\simeq \mathrm{Hom}(\pi_1(\cs),\Su)/\mathrm{gauge}$, where $\mathrm{Hom}(\pi_1(\cs),\Su)$ is the set of group homomorphisms from $\pi_1(\cs)$ to $\Su$ given by the loop holonomies.}. 

The Riemann surface $\cs$ considered here is generally made by gluing a number of 4-holed spheres. On each 4-holed sphere, the flat connection is dual to a constant curvature tetrahedron geometry (\cite{Hal.M.Haggard2015}, reviewed in Section \ref{tetra}). The (curved) tetrahedron closure condition is imposed by the flatness of connections. So the new LQG phase space and twisted geometry in this paper are already at the level of gauge invariant twisted geometries. 

\begin{figure}
	\centering\includegraphics[width=2.5in]{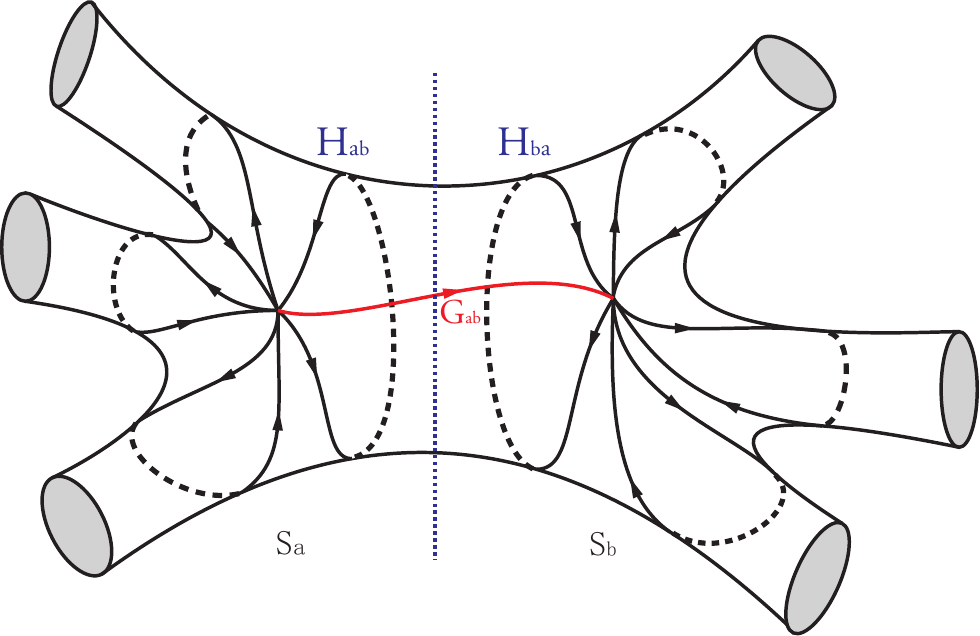}
	\caption{Two 4-holed spheres $S_a$ and $S_b$ glued together. }\label{fig:twe}
\end{figure}

When the 4-holed spheres are glued to form $\cs$, we define a longitude holonomy $G_{ab}$ of flat connection, which travels from one 4-holed sphere to another (see FIG.\ref{fig:twe}). It is shown in Section \ref{G4S} that $G_{ab}$ can be written in the following form
\be
G_{ab}=M_{ab}e^{\xi_{ab}\tau_3}M_{ba}^{-1},\label{keyformula}
\ee
In its geometrical interpretation, $M_{ab},M_{ba}\in SU(2)$ rotate $\hat{z}=(0,0,1)$ to the unit normals of a pair of glued faces from two tetrahedra, by the relation between flat connection on 4-holed spheres and constant curvature tetrahedra. The above form of $G_{ab}$ closely resembles the geometrical interpretation of the link holonomy in the usual twisted geometry. Here $\xi_{ab}$ plays the role of the twisted angle. The above formula suggests the close relation between SU(2) flat connection on Riemann surface and the generalized twisted geometry with cosmological constant, provided that the tetrahedra are of constant curvature in the present context.
 
Indeed the relation is made precise in Section \ref{RTG}, where we also identify the triangle area from the flat connection, and point out that a generalized area-matching condition is satisfied here. 

In Section \ref{sec:level3}, we derive the relation between flat connection variables on Riemann surface and 3D geometrical variables of interest in LQG, in particular, fluxes and holonomies of Ashtekar-Barbero connection. Firstly, We show in Section \ref{EFLUX} that in presence of cosmological constant, the usual LQG flux is naturally replaced by an exponentiated flux, which is the proper variable for a curved triangle face. The exponentiated flux is identified to the meridian holonomy $H_{ab}$ of the flat connection on the Riemann surface. 

Secondly, we shown in Section \ref{TAEC} that the longitude holonomy $G_{ab}$ on Riemann surface is identified with the LQG holonomy $G_{AB}$ of Ashtekar-Barbero connection along a link traveling from the interior of one tetrahedron to another. The Ashtekar-Barbero connection $A_\a^i=\G_\a^i+\g k_{\a}^i$ contains the extrinsic curvature $k_{\a}^i$. The 4D hyper-dihedral angle $\Theta_{AB}$ between a pair of neighboring tetrahedra, as the discrete version of $k_{\a}^i$, turns out to be proportional to the twist angle $\xi_{ab}$ in Eq.(\ref{keyformula}), i.e. $\xi_{ab}=\g\Theta_{AB}$ where $\g$ is the Barbero-Immirzi parameter.

As an consequence from the curvature of tetrahedroa and tetrahedron faces, the area-normal description of each constant curvature tetrahedron involves a choice of base point among the tetrahedron vertices, where the face normals are located. Different choices of tetrahedra's base points lead that the path of $G_{AB}$ is non-unique from one tetrahedron to another, in contrast to the situation of flat tetrahedra. This subtlety is discussed in Section \ref{MGCBP}. It turns out that different choices of paths of the LQG holonomy $G_{AB}$ are in 1-to-1 correspondence to the paths of $G_{ab}$ of the flat connection on the Riemann surface.

In Section \ref{esc:level4}, we discuss the symplectic structure of the new LQG phase space $\cm_{flat}(\cs,\Su)$ from SU(2) flat connection on Riemann surface. We show that the natural symplectic structure $\O$ on $\cm_{flat}(\cs,\Su)$ (derived from Chern-Simons theory) results in that the triangle area and twist angle $\xi_{ab}$ are the canonical conjugate variables\footnote{They relate to the famous Fenchel-Nielsen coordinates of flat connections, see Appendix \ref{FNcoo} for explanation.}, resembling the symplectic structure of twisted geometry. Combining the canonical variables of flat connection on 4-holed sphere dual to the curved tetrahedron in \cite{Hal.M.Haggard2015}, it shows the symplectic structure $\O$ on $\cm_{flat}(\cs,\Su)$ generalizes from the usual LQG to the situation with cosmological constant $\L$. As the limit of $\L\to0$, twisted geometry variables with flat tetrahedra and the symplectic structure are recovered from the symplectic coordinates of $\cm_{flat}(\cs,\Su)$ and their geometrical interpretations. 

Finally, we carry out the quantization of $\cm_{flat}(\cs,\Su)$ in Section \ref{QUAN}. The discussion here mainly focuses on the quantization of the symplectic coordinates which correspond to twisted geometry variables\footnote{More complete discussions on quantizing $\cm_{flat}(\cs,\Su)$ are given in e.g. \cite{Alekseev:1994pa,Elitzur:1989nr}.}. Quantizing the twisted geometry variables is shown to be the same as a quantum torus. Both the triangle area and twisted angle are quantized and have discrete spectra. The cosmological constant gives a cut-off to the area spectrum. Given the relation between twist angle $\xi_{ab}$ and extrinsic curvature, the discreteness of $\xi_{ab}$ leads to the discreteness of the hyper-dihedral angle $\Theta_{AB}$ at the quantum level. It might relate to the discreteness of time in LQG, as proposed in \cite{Rovelli:2015fwa}.

\section{\label{sec:level2}Flat Connection on Riemann Surface and Twisted Geometry }
\subsection{From Graph to Riemann Surface}\label{GRS}

As the first step to build a bridge between twisted geometries on graphs and generalized twisted geometries on Riemann surfaces, we define a bijection between graphs and Riemann surfaces. 

A graph contains a collation of links and vertices, while vertices are the end-points of links. Naively one can define a map $\rho$ from the set of graphs to the set of Riemann surfaces as follows: Firstly we relate each n-valent vertex to a 2D-sphere with n holes. Then we relate each link to a cylinder connecting a pair of holes on different n-holed spheres. By doing so we promote a graph to a surface, which defines $\rho$. However, the map $\rho$ is surjective but not injective.

The Riemann surface made by two n-holed spheres connected by a cylinder is topologically equivalent to a (2n-2)-holed sphere which means, its pre-image in $\rho$ may be either a graph with a single (2n-2)-valence vertex or a graph with two n-valence vertices connected by one link. So $\rho^{-1}$ is not single-valued.

\begin{figure}
	\centering\includegraphics{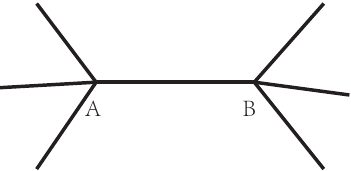}
	\caption{An example of graph. Two 4-valence vertices $A$ and $B$ are connected by a link.}\label{fig:spin}
\end{figure}

\begin{figure}
	\centering\includegraphics[width=1.5in]{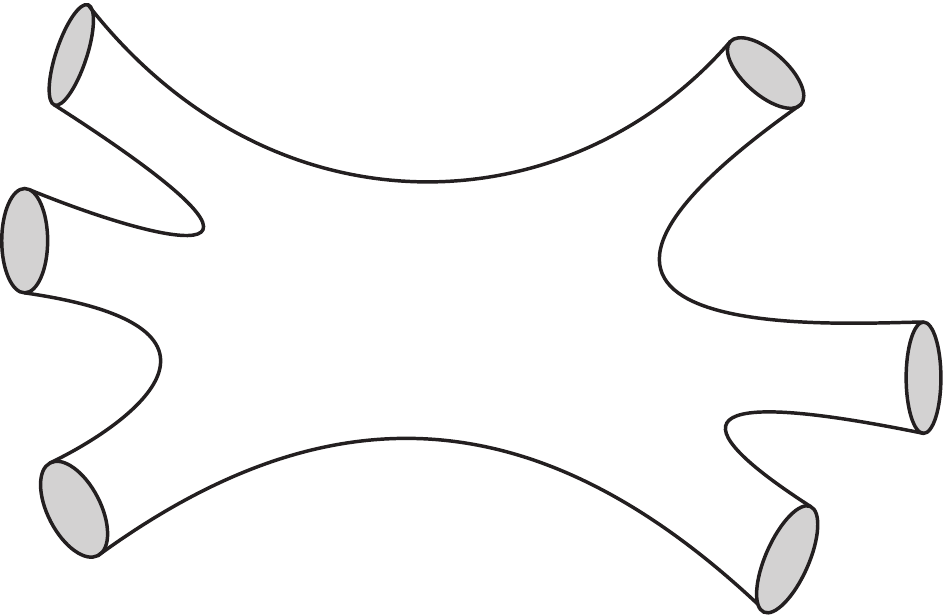}
	\caption{The Riemann surface made by connecting two 4-holed spheres.}\label{fig:twess}
\end{figure}

For example, assuming we have a graph made by a pair of 4-valence vertices connected by a link as it is showed in FIG.\ref{fig:spin}, $\rho$ maps this graph to a Riemann surface as FIG.\ref{fig:twess}.

\begin{figure}
	\centering\includegraphics[width=1in]{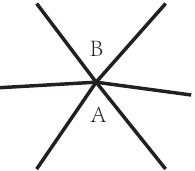}
	\caption{An alternative graph relating to FIG.\ref{fig:twess} by $\rho$}\label{fig:spin1}
\end{figure}

However, since FIG.\ref{fig:twess} is topologically equivalent to a 6-holed sphere, the graph contains a single 6-valence vertex like FIG.\ref{fig:spin1} also maps to FIG.\ref{fig:twess} by $\rho$. So the pre-image of FIG.\ref{fig:twess} is nonunique. $\rho$ is not a bijection. 

But we can construct a map $\hat{\rho}$ between graphs and Riemann surfaces as a bijection by decorating Riemann surfaces. Namely $\hat{\rho}$ is a map between the set of graphs and the set of decorated Riemann surfaces. More concretely, the idea of constructing $\hat{\rho}$ is nearly the same as $\rho$, except that $\hat{\rho}$ relates a link to a cylinder decorated with a meridian, and relates a n-valence vertex to a n-holed sphere decorated with a base point located on it. The identical Riemann surfaces with different decorations (meridians and base points) are understood as different decorated Riemann surfaces.  

As for the map $\hat{\rho}^{-1}$, the graph is uniquely recovered by connecting all base points on decorated Riemann surface with links, under the condition that each decorated meridian should only intersect one link.

\begin{figure}
	\centering
	\begin{minipage}[t]{0.45\linewidth}
		\centering
		\includegraphics[width=1.3in]{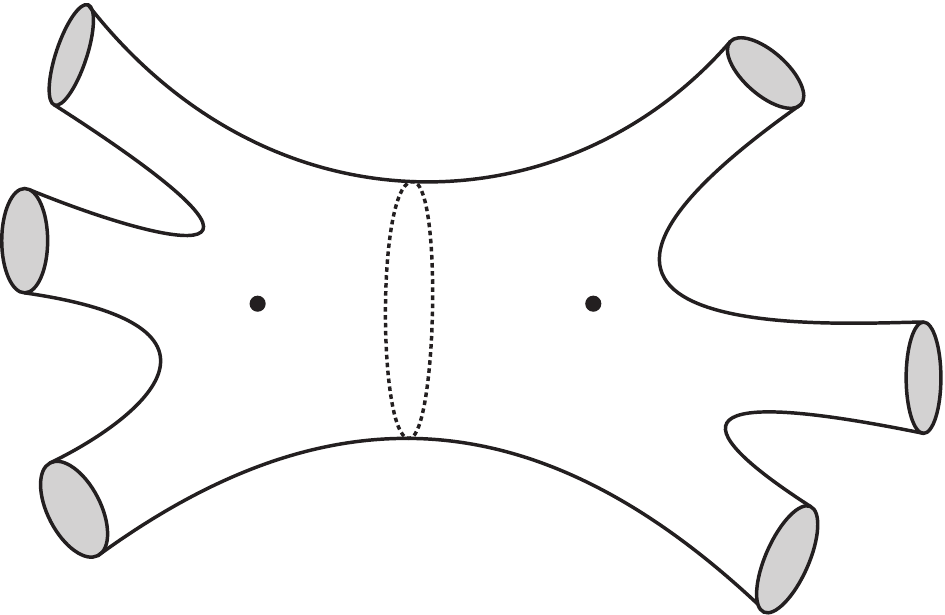}
		\caption{A Riemann surface decorated by two base points and a meridian.}\label{fig:twesd}
	\end{minipage}\hfill\begin{minipage}[t]{0.45\linewidth}
		\centering\includegraphics[width=1.3in]{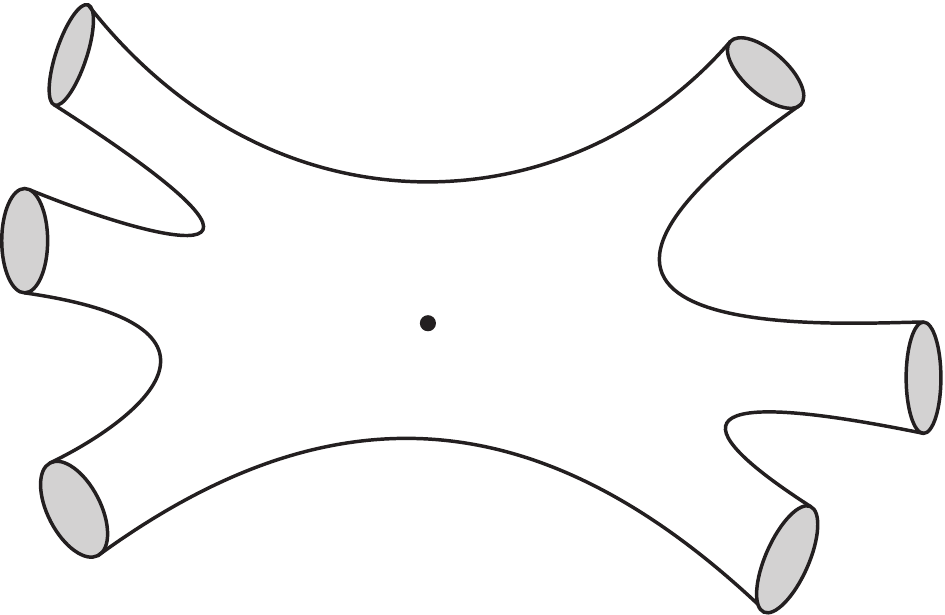}
		\caption{A Riemann surface decorated by a base point.}\label{fig:twesd2}
	\end{minipage}
\end{figure}

So $\hat{\rho}$ maps FIG.\ref{fig:spin} to a decorated Riemann surface FIG.\ref{fig:twesd} and maps FIG.\ref{fig:spin1} to a different decorated surface FIG.\ref{fig:twesd2}. It eliminates the ambiguity we mentioned before by distinguishing different decorations.

\begin{figure}
	\centering
	\begin{minipage}[t]{0.45\linewidth}
	\centering
	\includegraphics[width=1.3in]{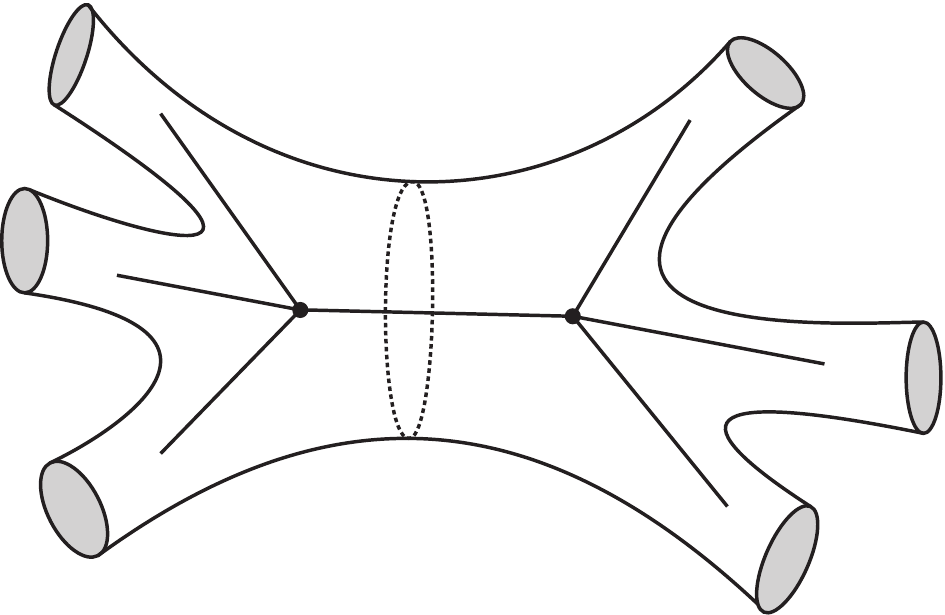}
	\caption{A graph recovered from Riemann surface FIG.\ref{fig:twesd}.}\label{fig:twesd1}
	\end{minipage}\hfill\begin{minipage}[t]{0.45\linewidth}
	\centering\includegraphics[width=1.3in]{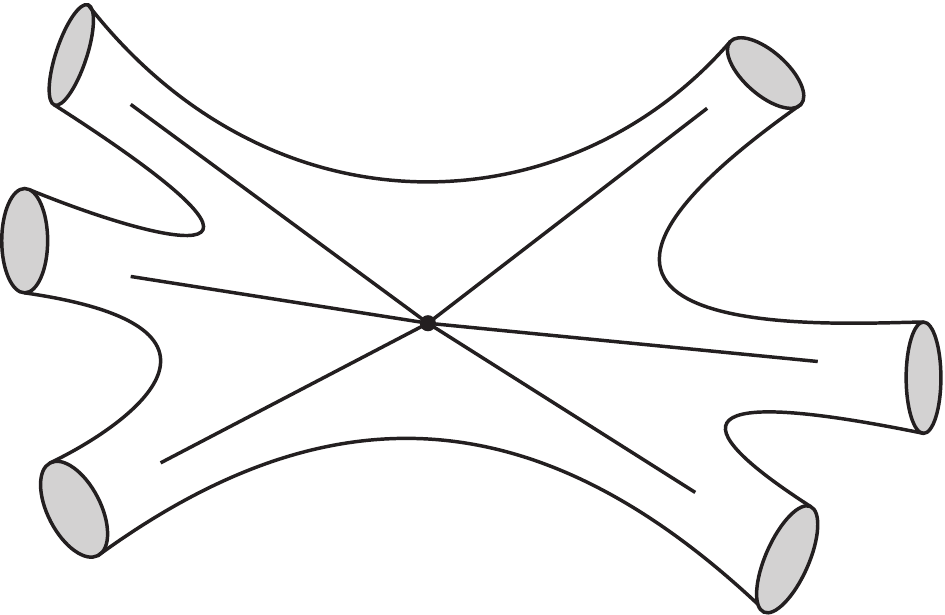}
	\caption{Graph recovered from Riemann surface FIG.\ref{fig:twesd2}.}\label{fig:twesd3}
	\end{minipage}
\end{figure}

The inverse of the map brings FIG.\ref{fig:twesd} and FIG.\ref{fig:twesd2} to FIG.\ref{fig:twesd1} and FIG.\ref{fig:twesd3} respectively without any ambiguity.

Although the naive map $\rho$ is not a bijection between graphs and Riemann surfaces, $\hat{\rho}$ is a bijection between graphs and decorated Riemann surfaces. Twisted geometries are defined on a cellular decomposition of 3D space. The cellular decomposition is dual to a graph. Because of the bijection between graphs and decorated Riemann surfaces. The cellular decomposition, where twisted geometries live, is equivalently understood as dual to a decorated Riemann surface. Therefore we are able to use decorated Riemann surface to study twisted geometries on discrete 3D space. 

Because all Riemann surfaces are decorated in the following discussion, in the rest part of the paper, we use the term ``Riemann surface'' to refer to the decorated Riemann surface.

\subsection{Flat Connection on Riemann Surface and Curved Tetrahedron}\label{tetra}

 
Given a 4-holed 2D-sphere denoted by $S_a$, according to \cite{Hal.M.Haggard2015,Haggard:2014xoa,2CCS,Haggard:2015yda}, there is a bijection between a flat $SU(2)$ connection defined on $S_a$ and a convex constant curvature tetrahedra geometry as far as the non-degenerate geometry is concerned.

The 4-holed sphere is considered as a decorated Riemann surface which has a base point. Denote by $H_i$ the $SU(2)$ holonomy along the loop which starts from the base point of the sphere, goes around the i-th hole\footnote{i, which is a number from 1 to 4, labels 4 holes on the sphere} on the sphere, and returns to the base point, we have
\begin{equation}\label{fun2}
\prod_iH_i=e.
\end{equation}

\begin{figure}
	\centering\includegraphics[width=2.5in]{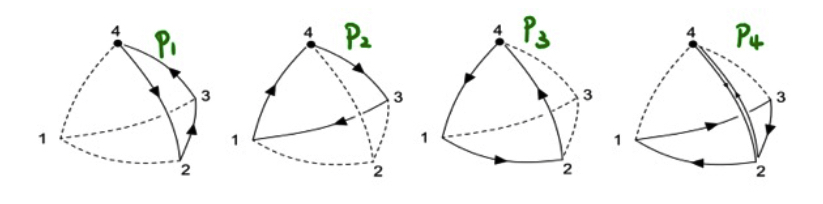}
	\caption{Path on tetrahedron, where point 4 is a base point and edge 2-4 is a special edge.}\label{fig:tetra}
\end{figure}

On the other hand, if we have a convex tetrahedra, by choosing a base point and a special edge we can define the closed paths along the boundary of each face. More specifically, if the base point is contained in the boundary of the face\footnote{Like the face $(1,2,4),(2,3,4)$ and $(1,2,4)$ in FIG\ref{fig:tetra}.}, the path will start from the base point, go around the face and end at the base point. If the base point is not contained in the boundary of the face\footnote{Like face $(1,2,3)$ in FIG.\ref{fig:tetra}.} the path will start from the base point, go along the special edge, move around the face, return back to base point through the special edge again. See FIG.\ref{fig:tetra}.

The spin-connection holonomies of those paths will obey the relation
\begin{equation}\label{fun1}
\prod_iU_{\partial f_i}=e,
\end{equation} 
which is identical to Eq.(\ref{fun2}) of the $SU(2)$ flat connection holonomies. It suggests an identification between the flat connection holonomies $H_i$ on $S_a$ and the spin-connection holonomies $U_{\partial f_i}$ on tetrahedron. 

Furthermore, \cite{2CCS} also shows that for the constant curvature tetrahedron whose faces are of vanishing extrinsic curvature, $U_{\partial f_i}$ relates to the area $A_i$ and normal $N_i$ of the face $f_i$ by
\begin{equation}\label{defr}
H_i=U_{\partial f_i}=\exp\left(\frac{\Lambda}{3}A_i N^i\tau_i\right)\in SU(2),
\end{equation}
where $N_i$ is the unit surface normal located at the base point, and $\tau_i=-\frac{i}{2}\sigma_i$ ($\sigma_i$ is the i-th Pauli matrix).

It turns out that \eqref{fun2} or \eqref{fun1} is the closure condition for the constant curvature tetrahedron. Namely given the data $A_i,N_i$ satisfying \eqref{fun2} or \eqref{fun1}, a unique convex tetrahedron of constant curvature $\Lambda$ can be reconstructed \cite{2CCS,Hal.M.Haggard2015}. As $\Lambda$ tends to be small, \eqref{fun2} reduces to the usual closure condition of the flat tetrahedron
\begin{equation}\label{clos}
\sum_i A_iN^i=0.
\end{equation}

\subsection{Gluing 4-holed Spheres}\label{G4S}

We expect that if we have a Riemann surface made by connecting two 4-holed spheres $S_a,S_b$ by a 2D-cylinder, the flat connection on this surface have a geometry interpretation as gluing two curved tetrahedra.


FIG.\ref{fig:twe} showed an example. The middle dash line implies the decoration-meridian. $H_{ab}$ stands for the holonomy around a hole on $S_a$. $H_{ba}$ goes around the corresponding hole on $S_b$ which is glued to the hole on $S_a$. $G_{ab}$ is the holonomy connecting two base points. We consider the flat SU(2) connection on the Riemann surface. The flatness implies 
\begin{equation}\label{gh}
H_{ab}=G_{ab}H_{ba}G_{ab}^{-1}.
\end{equation}
all $H_{ab},G_{ab},H_{ba}$ belong to SU(2).

$H_{ab}$ can be digonalized as
\begin{equation}\label{dia}
H_{ab}=M_{ab}\left( \begin{array}{cc}
x_{ab} & 0\\
0 & x_{ab}^{-1}
\end{array}\right) M_{ab}^{-1}
\end{equation}
Similarly for $H_{ba}$
\begin{equation}\label{dia1}
H_{ba}=M_{ba}\left( \begin{array}{cc}
x_{ba} & 0\\
0 & x_{ba}^{-1}
\end{array}\right) M_{ba}^{-1}
\end{equation}
Both of the matrices $M_{ab},M_{ba}$ belong to SU(2).

As a consequence of Eq.(\ref{gh}), the eigenvalues of $H_{ab}$ and $H_{ba}$ are identical 
\begin{equation}\label{identifyx}
x_{ab}=x_{ba}.
\end{equation}
Inserting Eqs.(\ref{dia}) and (\ref{dia1}) into \eqref{gh} shows that the combination $M_{ab}^{-1}G_{ab}M_{ba}$ commutes with $\mathrm{diag}(x_{ab},x_{ab}^{-1})$. It implies $M_{ab}^{-1}G_{ab}M_{ba}$ is diagonal
\begin{eqnarray}\label{dia2}
M_{ab}^{-1}G_{ab}M_{ba}=&\left( \begin{array}{cc}
e^{-i\xi_{ab}/2} & 0\\
0 & e^{i\xi_{ab}/2}
\end{array}\right),
\end{eqnarray}
which means
\begin{equation}\label{dia3}
G_{ab}=M_{ab}e^{\xi_{ab}\tau_3}M_{ba}^{-1}.
\end{equation}

Here we note that the parameter $\xi_{ab}$ is not uniquely determined by $G_{ab}$. In Eqs.(\ref{dia}) and (\ref{dia1}), $x_{ab}$ is invariant under the ``gauge transformation'' $M_{ab}\to M_{ab}e^{\alpha^\prime\tau_3}$ and $M_{ba}\to M_{ba}e^{\alpha^{\prime\prime}\tau_3}$. As a result, there is a gauge parameter $\alpha=\alpha^\prime-\alpha^{\prime\prime}$ appearing in equation \eqref{dia2} so that $\xi_{ab}\to\xi_{ab}-\alpha$. Eq.\eqref{dia3} becomes
\begin{equation}\label{dia4}
G_{ab}=M_{ab}e^{(\xi_{ab}-{\alpha})\tau_3}M_{ba}^{-1}.
\end{equation}

It turns out that $\ln x_{ab}$ and $\xi_{ab}$ are a pair of symplectic coordinates of SU(2) flat connections on Riemann surface. The freedom of $\alpha$ corresponds to the freedom in choosing the $\xi_{ab}=0$ in the coordinate system. The details are given in Appendix \ref{FNcoo}.

\subsection{Relation with Twisted Geometry }\label{RTG}

The equation \eqref{dia3}, coming from the SU(2) flat connection on Riemann surface, suggests a relation with the twisted geometry in 3-dimensions.

As it is introduced in \cite{LaurentFreidel2010}, the twisted geometry defined on a graph $\G$ has the phase space $S_2\otimes S_2\otimes T^*S_1$ on each link. The phase space can be parametrized by the collection of variables $(N,\tilde{N},j, \xi)$. $N=n\tau_3n^{-1}=N^i\t_i$ and $\tilde{N}=\tilde{n}\tau_3\tilde{n}^{-1}=\tilde{N}^i\t_i$ indicate the unit normals $N^i,\tilde{N}^i$ of the 2-face dual to the link. $N^i,\tilde{N}^i$ associates to the two ends of the link. $n,\tilde{n}$ are the rotations transforming $\hat{z}=(0,0,1)$ to the vectors $N^i,\tilde{N}^i$. $j$ is the area of the 2-face. The twist angle $\xi$ relates to the link holonomy $g$ by 
\begin{equation}\label{twi1}
g=ne^{\xi\tau_3}\tilde{n}^{-1}.
\end{equation} 
which shares the similarity with Eq.(\ref{dia3})

\begin{figure}
		\centering\includegraphics{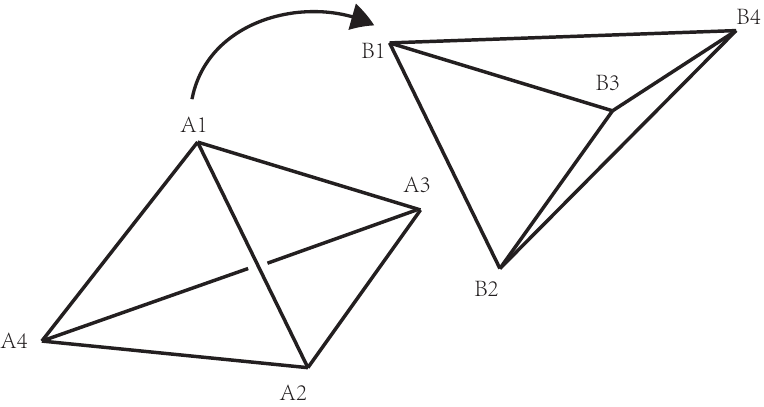}
		\caption{Gluing topologically two tetrahedra through the interface $(1,2,3)$. The topological gluing doesn't necessarily identify the geometry of the common face.}\label{fig:1}
	\end{figure}

The SU(2) flat connection on each of the pair of 4-holed spheres $S_a,S_b$ relates to the geometry of a constant curvature tetrahedron. Gluing $S_a,S_b$ and obtaining the Riemann surface FIG.\ref{fig:twe} suggest the topological gluing of two tetrahedra through a common face as FIG.\ref{fig:1}. Eq.(\ref{defr}) relates $H_{ab}$ to the geometry of a face of the tetrahedron associated to $S_a$. Then Eq.(\ref{dia}) implies that
\be
x_{ab}=\exp\lt(-\frac{i\L}{6}A_{ab}\rt),\ \ N^i_{ab}\t_i=M_{ab}\t_3 M_{ab}^{-1}
\ee
where $A_{ab}$ is the area of the face. $N^i_{ab}$ is the unit normal of the face located at the tetrahedron base point. The SU(2) matrix $M_{ab}$ is thus the rotation transforming $\hat{z}=(0,0,1)$ to $\vec{N}_{ab}$, playing precisely the same role as $n$ in twisted geometry. A similar interpretation is valid for $H_{ba}$. $M_{ba}$ plays the same role as $\tilde{n}$. The angles $\a',\a''$ are the freedom of rotations in the plane perpendicular to $\vec{N}_{ab},\vec{N}_{ba}$  As a result, the relation Eq.(\ref{dia3}) of $G_{ab}$ on Riemann surface resembles the twisted geometry equation (\ref{twi1}). $G_{ab}$ relates the two unit normals $\vec{N}_{ab},\vec{N}_{ba}$ of the gluing interface, where $\vec{N}_{ab}$ ($\vec{N}_{ba}$) is located at the base point of the left (right) tetrahedron. $G_{ab}$ again presents a twist angle $\xi_{ab}$ between $\vec{N}_{ab},\vec{N}_{ba}$, in the same way as the twisted geometry.

The resemblance between twisted geometry and SU(2) flat connection on Riemann surface may be summarized by the following:
\begin{equation}\label{trans}
\begin{split}
&n\leftrightarrow M_{ab},\ \tilde{n}\leftrightarrow M_{ba},\\
&j\leftrightarrow \ln x_{ab},\ \xi\leftrightarrow \xi_{ab}.
\end{split}
\end{equation}
It suggests that the twisted geometry can be generalized to the situation with constant curvature tetrahedron. The generalized twisted geometry relates naturally to the SU(2) flat connection on Riemann surface. The twist angle $\xi$ in the usual twisted geometry has been interpreted as the extrinsic curvature of the spatial slice, when $g$ is the holonomy of the Ashtekar-Barbero connection along the link  \cite{CarloRovelli2010,Langvik:2016hxn}. A similar interpretation can be obtain for $\xi_{ab}$ from $G_{ab}$ on Riemann surface, which is discussed in the next section. 

The identification $x_{ab}=x_{ba}$ is a generalization of the area-matching condition in twisted geometry. There is a key subtlety in comparing $x_{ab}=x_{ba}$ and the usual area-matching. $x_{ab}$ relates $A_{ab}$ via an exponential. The periodicity restricts $A_{ab}\in[0,\frac{12\pi}{|\L|}]$. However there is no restriction to guarantee that the constant curvatures $\L$ are the same from the pair of tetrahedra. It may happen that the first tetrahedron corresponding to $S_a$ is spherical ($\L>0$), while the second tetrahedron corresponding to $S_b$ is hyperbolic ($\L<0$). Due to this subtlety, $x_{ab}=x_{ba}$ doesn't restrictively implies the area matching $A_{ab}=A_{ba}$, but rather contains an ambiguity. More precisely, it implies that
\be
A_{ab}=A_{ba}\ \ \text{or}\ \ \frac{|\L|}{3}A_{ab}=2\pi-\frac{|\L|}{3}A_{ba}.\label{ambig}
\ee
The source of this ambiguity is the proper interpretation of tetrahedron face area from the closure condition Eq.\Ref{fun1}. The details can be found in \cite{2CCS,Hal.M.Haggard2015} \footnote{In the case of small area $A_{ab},A_{ba}\ll |\L|^{-1}$, the second possibility cannot hold, which resolves the ambiguity.}. 

At the quantum level, in the spinfoam model with cosmological constant \cite{Haggard:2014xoa}, $x_{ab}$ relates to the spins $j_{ab}\in\mathbb{N}/2$ by $x_{ab}=\exp (\frac{2\pi i}{k}j_{ab})$, where the integer $k$ is the Chern-Simons level. So $x_{ab}=x_{ba}$ implies $j_{ab}=j_{ba}$ which is the same identity as the spin-network. The above ambiguity of area-matching comes from the ambiguity in interpreting the spins $j_{ab}$ as areas at the level of constant curvature tetrahedron. However the ambiguity Eq.(\ref{ambig}) is resolved dynamically in the semiclassical limit of spinfoam amplitude, which exhibits more constraints than the twisted geometry, and implies the gluing of tetrahedra with both area and shape matchings of their faces \cite{Haggard:2014xoa}.   

In the present paper, we work at the level of twisted geometry, and admit that there is an ambiguity Eq.(\ref{ambig}) of area-matching when the twisted geometry is generalized to curved tetrahedra. The ``generalized area-matching condition'' is $x_{ab}=x_{ba}$ in the present context.

As another important remark, the SU(2) flat connection on Riemann surface automatically implies the closure condition Eq.(\ref{fun1}). Therefore the twisted geometry obtained from SU(2) flat connection on Riemann surface is at the level of the gauge invariant twisted geometry (\`a la \cite{LaurentFreidel2010}), in which the tetrahedron closure condition has been implemented.

\section{\label{sec:level3}Geometric interpretation of Twisted Geometry}

	In \cite{CarloRovelli2010} and \cite{Langvik:2016hxn}, a relation between twist angle and extrinsic curvature has been established for the usual twisted geometry. The twist angle $\xi$ can be interpreted as $\gamma \Theta$, where $\Theta$ is the hyper-dihedral (boost) angle between two tetrahedra, and $\g$ is the Barbero-Immirzi parameter. The hyper-dihedral angle $\Theta$ is a discrete version of extrinsic curvature as $k_{\a\b}$. The flux variable $X$ in LQG is interpreted as $jN^i$ where $j$ is the area and $N^i$ is the normal of the face. 
	
	In this section, we show that the geometric interpretation of twisted geometry on graph remains valid in the generalized twisted geometry from flat SU(2) connection on Riemann surface. 
	
	One of the key difference between a geometry made piecewise by curved tetrahedra and a traditional Regge geometry is that, instead of the flatness inside each tetrahedra for the usual Regge case, it is curved inside each of the tetrahedra in our case. Just as the difference between special relativity and general relativity, we can no longer directly compare or inner product vectors at different points in space. Comparing vectors at different points involves a parallel transportation. 
	
	Consider in 4D spacetime, a 3D spatial slice triangulated by constant curvature tetrahedra. For each tetrahedron, instead of defining a general space-like normal vector for its surfaces, we have to specify the space-like normal vector of each surfaces at a certain base point of the surface. Similarly, the 4D normal of the tetrahedron should be defined at the base point of the tetrahedron. 
	
	\subsection{Exponentiated Flux}\label{EFLUX}

	The flux $X$ of a face $f$ is used to be defined as $\int_f e \wedge e$ in the usual context of LQG or twisted geometry. However it has been suggested in \cite{Hal.M.Haggard2015,Dittrich:2014wpa}, that in the presence of cosmological constant or constant curvature tetrahedron, $X$ should be replaced by a suitable version of exponentiated flux.  
	
Indeed we consider the 3D spin connection $\G_\a^i$ determined by the triad $e_\a^i$, and define the holonomy of $\G_\a^i$ along the boundary of each tetrahedron face\footnote{If the base point is not contained in the face, then the holonomy will use the special edge to connect the base point and the face as it is showed in the last picture in FIG.\ref{fig:tetra}.}. By non-abelian stokes theorem, we have
	\begin{equation}\label{mhol2}
	\begin{split}
	U_{\partial f}=&\mathcal{P} \exp\lt(\oint_{\partial f} \G\rt)\\
	=&\mathcal{P} \exp \lt(\int_f U(x) R U^{-1}(x) \rt), 
	\end{split}
	\end{equation}
where $\mathcal{P}$ stands for path ordering in the first line and surface ordering in second line. $R$ is the curvature obtained from $\G_\a^i$. A path system has been chosen on $f$ such that for each point $x\in f$, there is a path $p_x$ connecting $x$ to the base point of $U_{\partial f}$. $U(x)$ is the parallel transportation by $\G$ along the path $p_x$.
	
	Within each tetrahedron, the geometry is of constant curvature, which implies $R=\frac{\Lambda}{3} e \wedge e$. Therefore
	\be\label{mhol}
	U_{\partial f}&=&\mathcal{P} \exp \left(\frac{\Lambda}{3}\int_f U(x) (e\wedge e) U^{-1}(x) \right)\\
	&=& \mathcal{P} \prod_{x}\exp \left[\frac{\Lambda}{3} U(x) \lt(\epsilon^{\a\b} e_\a  e_\b \delta A(x)\rt) U^{-1}(x) \right].\nonumber
	\ee 
We have discretized the integral in the above. Each point $x$ is contained in a plaquette whose area is $\delta A(x)$. Here $U(x)$ parallel transport $\epsilon^{\a\b} e_\a  e_\b$ to the base point of $U_{\partial f}$ along $p_x$. Moreover $f$, as a face of constant curvature tetrahedron, is a flatly embedded surface (vanishing extrinsic curvature) in 3d constant curvature space. $f$ being a flatly embedded surface implies that the quantity $\frac{\Lambda}{3} U(x) \lt(\epsilon^{\a\b} e_\a  e_\b\rt) U^{-1}(x)$ (located at the base point) is independent of $x$ \cite{Haggard:2014xoa}.

At the base point, the wedge product of two orthonormal frame vectors along the interface gives the surface normal $\vec{N}$ which in $SU(2)$ representation is $N^i\tau_i$. So equation \eqref{mhol} reduces to
	\begin{equation}\label{mhol1}
	U_{\partial f}=\exp\left(\frac{\Lambda}{3} A N^i \tau_i\right)
	\end{equation} 
	where $A$ is the area of the surface. In the usual twisted geometry the flux variable is given by $\vec{X}=A \vec{N}$ where $\vec{N}$ is the normal of a flat tetrahedron face. The new variable $U_{\partial f}$, which is natural in the present context of curved tetrahedron, is manifestly an exponentiated flux variable, with $\vec{N}$ being the face normal at the base point. 
	
	This actually proves the equation \eqref{defr}. By the correspondence between constant curvature tetrahedron and SU(2) flat connection on 4-holed sphere \cite{Hal.M.Haggard2015}, $U_{\partial f}$ is identified with the holonomy $H_{ab}$ of flat connection.
	
	For a pair of tetrahedra topologically glued as in FIG.\ref{fig:1}, their interface $f=(1,2,3)$ has two exponentiated fluxes $U_{\partial f}$ and $\tilde{U}_{\partial f}$ associated to two different tetrahedra. They relate respectively to two different normals $\vec{N}_{ab}$ and $\vec{N}_{ba}$ located at the base points of tetrahedra A and B.
	
	In the reconstruction of constant curvature tetrahedron geometry from flat SU(2) connection on $S_a$ or $S_b$, all the resulting face normals of tetrahedron A and tetrahedron B are located at their base points respectively.

\subsection{Twist Angle and Extrinsic Curvature}\label{TAEC}

Take Figure.\ref{fig:1} as an example of gluing two tetrahedra, tetrahedron A is gluing with tetrahedron B through the interface labeled by $(1,2,3)$, which means the point $A1$ identifies with $B1$ after the gluing, so do the points $A2$, $A3$ and $B2$, $B3$.
	
We firstly consider a simple case: We set tetrahedron A and tetrahedron B share the same base point at $A1$ as well as $B1$. In the following, this point is often mentioned as point $1$ for abbreviation, and the interface $(1,2,3)$ will be denoted as $f_{AB}$.

In order to discuss the extrinsic curvature $k$ and its relation with the twist angle $\xi_{ab}$, we consider a Regge geometry on the spatial slice, which is made by tetrahedra with constant curvature $\L$. We zoom into a pair of glued tetrahedra as FIG.(\ref{fig:1}). In the present situation, both of the 4D (timelike) normals of two tetrahedra are defined at the common base point $1$, so does the derivative of 4D normals. Although in the usual Regge geometry the extrinsic curvature is smeared on the entire interface \cite{CarloRovelli2010}, it is reasonable now to regulate the smeared extrinsic curvature in a neighborhood at the base point, since the extrinsic curvature is the derivative of 4D normal. It is also consistent with the semiclassical geometry emerged from spinfoam \cite{Haggard:2014xoa}, in which the discrete extrinsic curvature is defined at the base point, instead of being defined at the common face. 
	
By the above argument, the discrete extrinsic curvature is given by
	\begin{equation}\label{eq:exs}
	k_{\a\b}(x)=\Theta_{AB} \int_{f_\epsilon}N_\a(x) N_\b(x)\delta^3(x,f_\epsilon(\sigma))d^2\sig,
	\end{equation} 
$f_\epsilon$ is the neighborhood of point $1$ on face $f_{AB}$. $N_\a(x)$ is the 3D normal vector field on $f_{AB}$. $N_\a$ is not a constant since the face is curved. $k_{\a\b}$ only has the component normal to $f_{AB}$ because it describes the change of 4D normals across $f_{AB}$. The plane where 4D normal rotates is orthogonal to $f_{AB}$. $\Theta_{AB}$ is the boost angle (hyper-dihedral angle) between the 4D-normals of the two tetrahedra \cite{CarloRovelli2010,Langvik:2016hxn}.
	
	\begin{figure}
		\centering\includegraphics[width=2in]{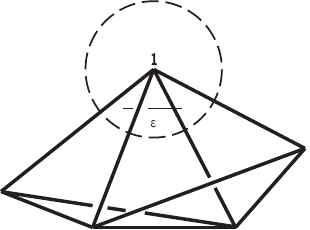}
		\caption{The dash circle stands for a small open sphere around point $1$. The link $\epsilon$ is at the vicinity of the base point of tetrahedra. }\label{fig:int}
	\end{figure}
	
We define a holonomy $G_{AB}$ of Ashtekar-Barbero connection $A^i_\a=\Gamma^i_\a+\gamma k^i_{\a}$ traveling within a small neighborhood at the common base point. $G_{AB}$ is along an infinitesimal link $\epsilon$ which intersect $f_{AB}$ transversely. 	
	\begin{equation}\label{eq:hol}
	G_{AB}=\mathcal{P}\exp\int_{\epsilon}d\epsilon^\a(\Gamma^i_\a+\gamma e^{\b i}k_{\a\b})\tau_i,
	\end{equation}
$e^{\a i}$ is a triad defined in the neighborhood.
	
At the vicinity of the base point, we can choose a smooth triad field $e^i_\a$ in tetrahedron A and extend smoothly to tetrahedron B. The 3D spin connection $\G_{\a}^i$ determined by $e^i_\a$ is a smooth field when crossing $f_{AB}$\footnote{The 3D curvature of $\G_{\a}^i$ is smooth except at each internal edge. The internal edge is the hinge of 3D deficit angle. }. Thus in $G_{AB}$, the contribution from the spin connection $\Gamma^i_\a$ is tiny since $\epsilon$ is infinitesimal. The main contribution comes from the extrinsic curvature. Combing \eqref{eq:exs}, we have
\be\label{eq:hols}
	G_{AB}&\simeq &\cp\exp\int_{f_\epsilon}\int_{\epsilon}\gamma e^{\b i} \Theta_{AB} N_\a N_\b\delta^3(x,f_\epsilon(\sigma)))\tau_i\, d\epsilon^\a d^2\sigma\nonumber\\
	&=&\exp\lt(\gamma \Theta_{AB} N^i\tau_i\rt)
\ee
where $N^i=e^{\b i} N_\b$ is located at the intersection between $\epsilon$ and $f_{\epsilon}$. As the limit $\epsilon\to 0$, the intersection approaches to the common base point (point 1). 
	
We are free to perform a gauge transformation at one end of $\epsilon$ on the tetrahedron B side. The gauge transformation is written as $V=M_{ab}(M_{ba}e^{\alpha\tau_3})^{-1}$ where the notions $M_{ab},M_{ba},\a$ are explained in a moment. After the gauge transformation $G_{AB}\rightarrow G_{AB}V$ is written as\footnote{The periodicity of $\gamma \Theta_{AB}$ on the exponential reflects the compact-ness of the space of SU(2) Ashtekar-Barbero connection. }
\begin{equation}\label{eq:holf}
	\begin{split}
	G_{AB}&=M_{ab}e^{\gamma \Theta_{AB} M_{ab}^{-1}N^i\tau_iM_{ab}}e^{-\alpha\tau_3}M_{ba}^{-1}\\
	&\simeq M_{ab}e^{(\gamma \Theta_{AB} -\alpha)\tau_3}M_{ba}^{-1}
	\end{split}
	\end{equation}
As the limit $\epsilon\to0$, $M_{ab}\in SU(2)$ has been set to be the rotation transforming $\hat{z}$ to $\vec{N}$. $\vec{N}\equiv \vec{N}_{ab}$ is understood as the normal of $f_{AB}$ at the base point of tetrahedron A. $M_{ba}\in SU(2)$ rotations the local frame in tetrahedron B, and rotation $\hat{z}$ to a new vector $\vec{N}_{ba}$, understood as the normal of $f_{AB}$ at the base point of tetrahedron B (as the limit $\epsilon\to0$). Namely $M_{ab},M_{ba}$ have the same geometrical meaning as the ones interpreted previously in Eq.(\ref{dia3}). The angle $\a$ is again the rotation freedom in the plane perpendicular to the face normal. 

Comparing Eqs.(\ref{eq:holf}) and (\ref{dia4}), we find the flat connection holonomy $G_{ab}$ on Riemann surface can be identified with the (infinitesimal) holonomy $G_{AB}$ of Ashtekar-Barbero connection in 3D space discretized by constant curvature tetrahedra. It relates the twist angle $\xi_{ab}$ to the hyper-dihedral boost angle $\Theta_{AB}$ by	
	\begin{equation}\label{eq:mean}
	\xi_{ab}=\gamma \Theta_{AB}.
	\end{equation}

	\subsection{More General Choices of base points}\label{MGCBP}

The path $\epsilon$ of $G_{AB}$ being infinitesimal is an artifact from assuming tetrahedra A and B to share the same base point. If two tetrahedra have different base points, $G_{AB}$ will be not infinitesimal. Consequently, we also need to take the different shapes of the path into consideration.
	
	In general, $G_{AB}$ is defined as the holonomy of Ashtekar-Barbero connection along the path connecting two base points of two tetrahedra. When two points coincide, the path of $G_{AB}$ is defined un-ambiguously as in previous section. However when two base points doesn't coincide, the path connecting them may be non-unique.
	
\begin{figure}
		\centering\includegraphics[width=2in]{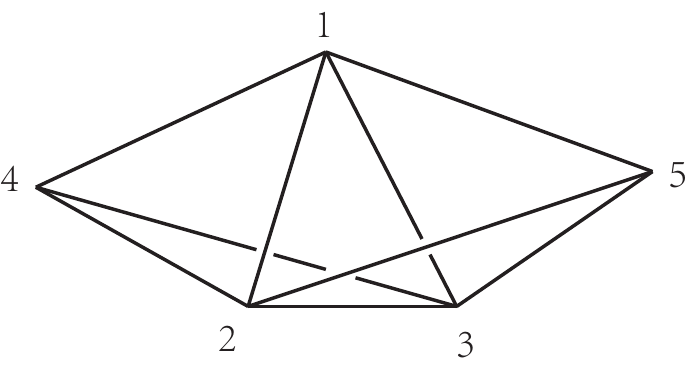}
		\caption{Two tetrahedra with their vertices labeled by numbers through 1 to 5.}\label{fig:patht1}
	\end{figure}

	Consider again two tetrahedra glued together. For convenient, we label the vertices by the number through 1 to 5 as it is showed in FIG.\ref{fig:patht1}. 
	
	As mentioned in Section.\ref{tetra}, the base point and special edge have to be specified on each tetrahedra so that the tetrahedron geometry relates to a flat connection on 4-holed sphere. Let's define point $4$ and point $5$ in FIG.\ref{fig:patht1} as the base point for two tetrahedra respectively and define edge $4-2$ and edge $5-1$ as the respective special edges.
	
\begin{figure}
		\centering
		\begin{minipage}[t]{0.45\linewidth}
			\centering
			\includegraphics[width=1.7in]{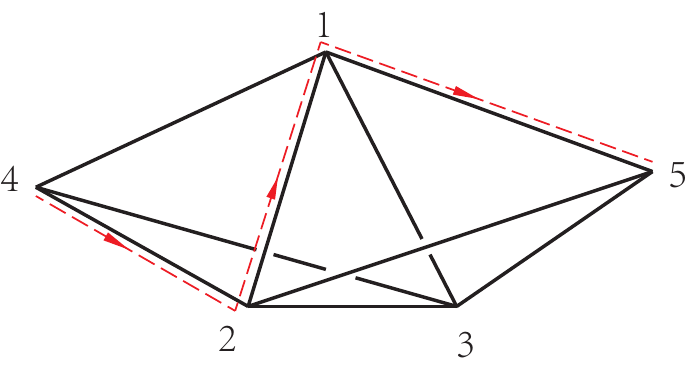}
			\caption{The path $4\to2\to1\to5$ which is labelled as $p_0$ }\label{fig:patht2}
		\end{minipage}\hfill\begin{minipage}[t]{0.45\linewidth}
		\centering
		\includegraphics[width=1.7in]{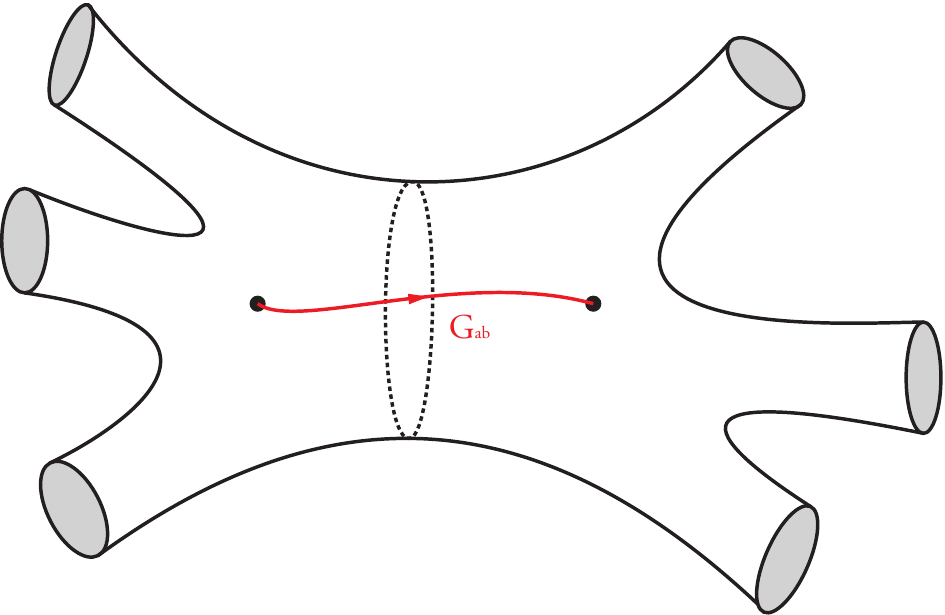}
		\caption{The path $\hat{p}_0$ which is the path corresponds to the path $p_0$ on a pair of tetrahedra.}\label{fig:pathr1}
		\end{minipage}
	\end{figure}

\begin{figure}
		\centering
		\begin{minipage}[t]{0.45\linewidth}
			\centering
			\includegraphics[width=1.7in]{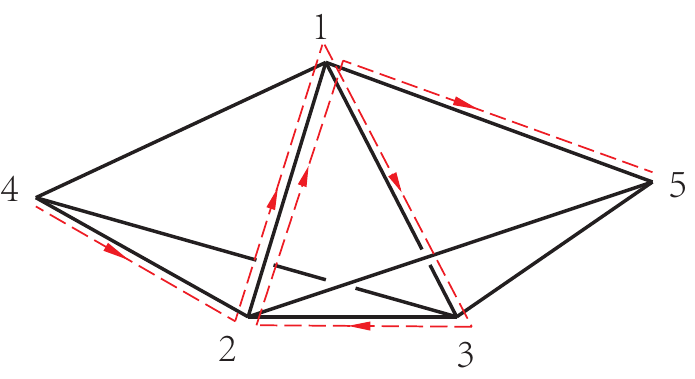}
			\caption{The path $p_{1}$ constructed by adding $p_0$ with one left-handed winding.}\label{fig:patht5}
		\end{minipage}\hfill\begin{minipage}[t]{0.45\linewidth}
			\centering
			\includegraphics[width=1.7in]{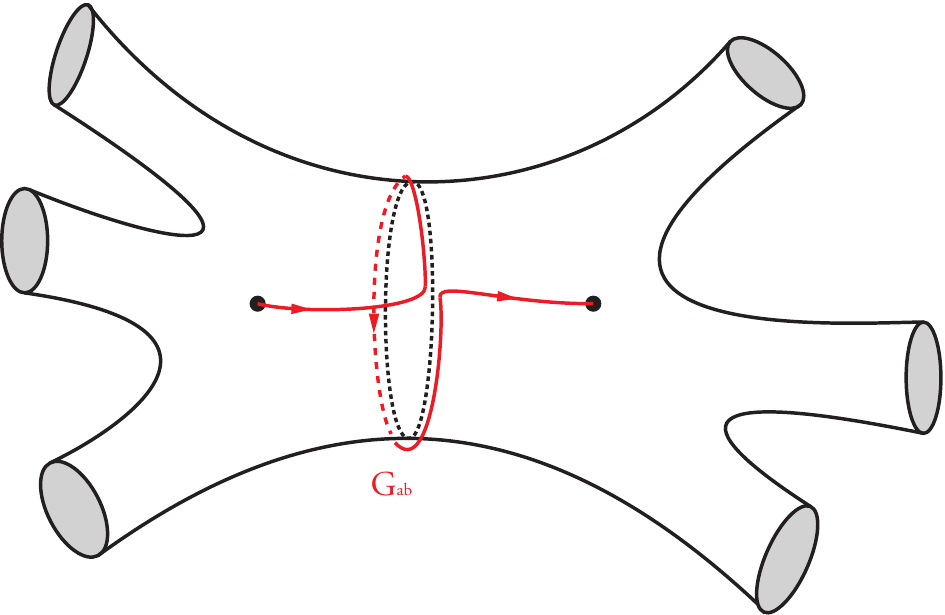}
			\caption{The path $\hat{p}_1$ made by adding $\hat{p}_0$ with a left-handed winding.}\label{fig:pathr3}
		\end{minipage}
	\end{figure}

\begin{figure}
		\centering
		\begin{minipage}[t]{0.45\linewidth}
		\centering
		\includegraphics[width=1.7in]{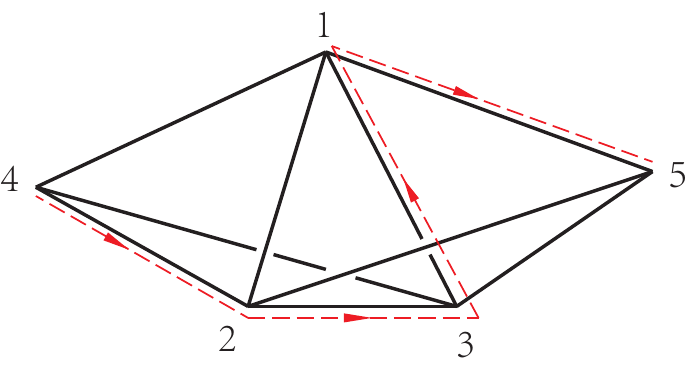}
		\caption{The path $4\to2\to3\to1\to5$ which is labeled as $p_{-1}$}\label{fig:patht3}
		\end{minipage}\hfill\begin{minipage}[t]{0.45\linewidth}
		\centering
		\includegraphics[width=1.7in]{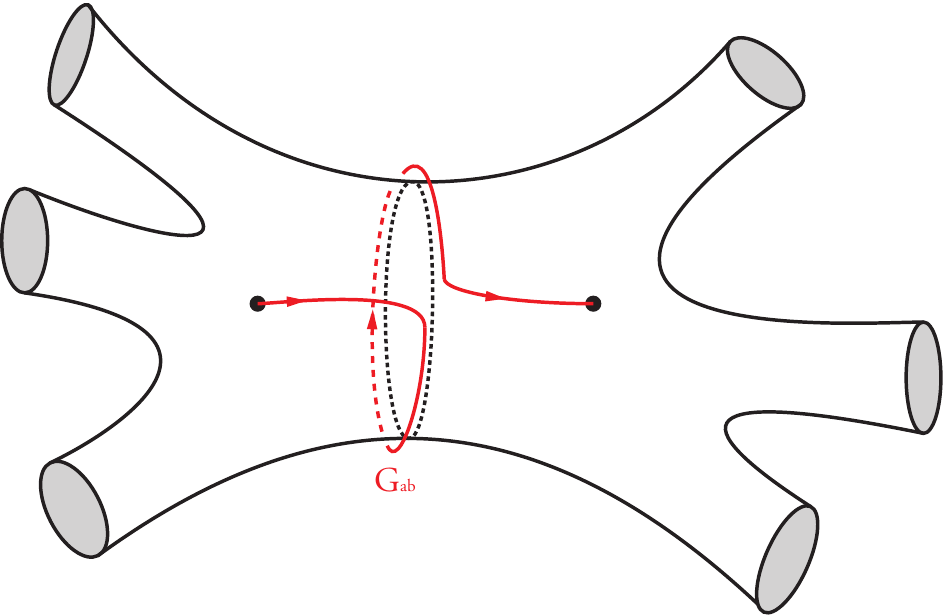}
		\caption{The path $\hat{p}_{-1}$ made by adding $\hat{p}_0$ with a right-handed winding.}\label{fig:pathr2}
		\end{minipage}
	\end{figure}

\begin{figure}
		\centering
		\begin{minipage}[t]{0.45\linewidth}
		\centering			\includegraphics[width=1.7in]{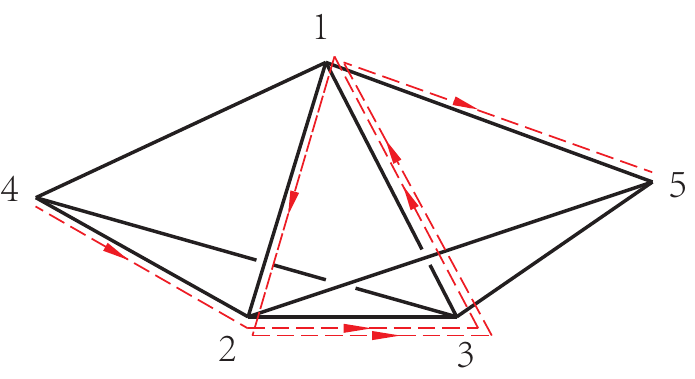}			\caption{The path $p_{-2}$ made by $p_0$ combining with two right-handed windings.}\label{fig:patht4}			
		\end{minipage}\hfill\begin{minipage}[t]{0.45\linewidth}
		\centering
		\includegraphics[width=1.7in]{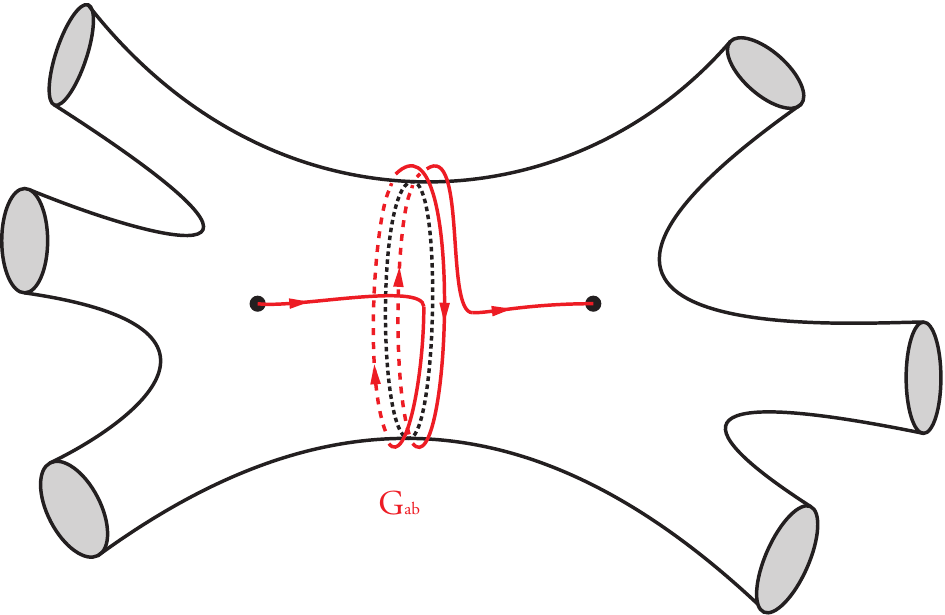}
		\caption{The path $\hat{p}_{-2}$ made by adding two right-handed windings to $\hat{p}_0$}\label{fig:pathr4}
		\end{minipage}
	\end{figure}

	We may choose the path of $G_{AB}$ to sequentially pass through $4\to2\to1\to5$ as it is showed in FIG.\ref{fig:patht2}\footnote{For convenience, we set the path of $G_{AB}$ always contain the special edges when the base points are not on the interface. It is consistent with the choice of path for $G_{ab}$ on Riemann surface shown in FIG.\ref{fig:tetra}.}. Let's denote this path as $p_0$. However, there is another path passing through $4\to2\to3\to1\to5$ as it is showed in FIG.\ref{fig:patht3}, denoted by $p_{-1}$. Clearly path $p_{-1}$ is just path $p_0$ plus an additional closed winding $1\to2\to3\to1$. Similarly we may add a closed winding $1\to3\to2\to1$ to make the path looks like FIG.\ref{fig:patht5}, denoted by $p_{1}$, or add two more closed windings $1\to2\to3\to1$ to $p_0$ to get a path like FIG.\ref{fig:patht4}, denoted by $p_{-2}$. We define the winding $1\to2\to3\to1$ as right-handed and $1\to3\to2\to1$ as left-handed. These examples indicate that (1) the path connecting the base points 4 and 5 is not unique, and (2) the paths connecting $4$ and $5$ can be classified by the number of windings along the boundary of interface, when the paths contain the special edges of two tetrahedra. We can make any path by adding right-handed or left-handed windings to $p_0$. We label the path between $4$ and $5$ by $p_i$, where $i\in\mathbb{Z}$ is the winding number.

The above paths $p_i$ are in 1-to-1 correspondence to the paths for $G_{ab}$ of the flat connection on Riemann surface. 

On a Riemann surface like FIG.\ref{fig:twesd1}, the holonomy $G_{ab}$ is along the path connecting two base points of the 4-holed spheres. The paths is again not unique. With the pair of base points fixed, the homotopy classes of the paths are again classified by the windings along the meridian\footnote{Within each homotopy class, the paths give the same $G_{ab}$ because the connection is flat.}. For instance, we may have a path connecting the base points as it is showed in FIG.\ref{fig:pathr1}, denoted by $\hat{p}_0$. We may draw some different paths $\hat{p}_i$ by adding $\hat{p}_0$ with right-handed windings $(i<0)$ or left-handed windings $(i>0)$ along the meridian. FIG.\ref{fig:pathr1}, FIG.\ref{fig:pathr3}, FIG.\ref{fig:pathr2} and FIG.\ref{fig:pathr4} are the examples of these paths.

	
As the arrangement of the pictures in previous page indicates, a bijection can be defined by relating $p_i$, being the path of $G_{AB}$ on tetrahedra, to $\hat{p}_i$ which is the path of holonomy $G_{ab}$ on Riemann surface.

However, no matter along which path the holonomy $G_{AB}$ on tetrahedra is defined, the common features are that (1) the path goes from one tetrahedron to the other, and (2) the path pass through at least one vertex of the interface triangle for at least once. So we are able to choose one vertex $P$ of the interface $f_{AB}$. The vicinity of $P$ is understood to contain the intersection between the path and the interface, as the path travels from one tetrahedron to the other. More precisely, we perform a regularization so that the path is not precisely along the edges of tetrahedra but is rather located slightly inside the tetrahedra. It has an infinitesimal distance $\epsilon$ from the tetrahedron edges. 

The (regularized) path of $G_{AB}$ can be divided into three segments. The 1st segment (with holonomy $G_{PB}$) connects the base point of tetrahedron B to the vicinity of the chosen interface vertex $P$, the 2nd segment is the infinitesimal link $\epsilon$ in the vicinity of $P$, and intersects transversely the interface $f_{AB}$, the same as the one in Section \ref{TAEC}. The 3rd segment (with holonomy $G_{AP}$) connects the vicinity of $P$ to the base point of tetrahedron A\footnote{When one of the base point is on the interface, we only need to divide the path of $G_{AB}$ into two segments, the link $\epsilon$ and the segment connecting to the other base point.}. As it is mentioned in the last paragraph, the 1st and 3rd segments of $G_{PB}$ and $G_{AP}$ are not precisely along the edges of tetrahedra, but slightly located inside the tetrahedron B and A. 

Denote the separation of $G_{AB}$ as
\begin{equation}\label{eq:G}
G_{AB}=G_{AP}G_{\epsilon}G_{PB}
\end{equation}
Recall $G_{AB}$ is the holonomy of Ashtekar-Barbero connection $A_\a^i=\G_\a^i+\g k_\a^i$. In the discretization, the extrinsic curvature Eq.(\ref{eq:exs}) is only located at the interface $f_{AB}$, and inside the neighborhood $f_\epsilon$ of the chosen vertex $P$ (So the hyper-dihedral angle $\Theta_{AB}$ is defined at $P$). So the extrinsic curvature $k_\a^i$ only contributes $G_{\epsilon}$, while $G_{AP},G_{PB}$ only receive the contribution from the spin connection $\G_\a^i$, since they are slightly inside the tetrahedra A and B. Thus $G_{AP}$ or $G_{PB}$ are parallel transports relating the reference frames at different points inside tetrahedron A or B. $G_{AP}^{-1}$ parallel transports the interface normal $\vec{N}_{ab}$ from the base point of tetrahedron A to the vicinity of $P$, and $G_{PB}$ parallel transports $\vec{N}_{ba}$ from the base point of tetrahedron B to the vicinity of $P$. 

Within the vicinity of $P$, $G_\epsilon$ has been computed in Eqs \eqref{eq:hols} and \eqref{eq:holf}. As a result, we obtain again 
\begin{equation}\label{eq:holf1}
	\begin{split}
	G_{AB}&=M_{ab}e^{(\gamma \Theta_{AB} -\alpha)\tau_3}M_{ba}^{-1},
	\end{split}
\end{equation}
Here $M_{ab},M_{ba}$ is different from the ones in \eqref{eq:holf} up to the additional parallel transports $G_{AP},G_{PB}$. But the new $M_{ab},M_{ba}$ have the right geometrical meaning as the ones interpreted previously in Eq.(\ref{dia3}). Namely $M_{ab}$ ($M_{ba}$) rotations $\hat{z}$ to the normal $\vec{N}_{ab}$ ($\vec{N}_{ba}$). The normals $\vec{N}_{ab},\vec{N}_{ba}$ of the interface $f_{AB}$ are located respectively at the base points of tetrahedra A and B. $\Theta_{AB}$ is the hyper-dihedral boost angle at $P$.

Comparing to Eq.\Ref{dia4}, identifying $G_{ab}$ to $G_{AB}$ relates the twist angle to the hyper-dihedral angle 
\be
\xi_{ab}=\gamma\Theta_{AB}.
\ee 



There is a useful remark: The above discussion starts from the flat connection on Riemann surface and proposes the 3D geometrical interpretation to the flat connections. However it may be helpful to consider a reverse logic: One may start from the 3D discrete geometry with constant curvature tetrahedra, and construct the holonomy $G_{AB}$ of Ashtekar-Barbero connection, as well as the exponential flux $U_{\partial f}$. The discussion in Sections \ref{TAEC} and \ref{MGCBP} shows Eq.(\ref{eq:holf1}) for $G_{AB}$, while the discussion in Section \ref{EFLUX} shows Eq.(\ref{mhol1}) for $U_{\partial f}$. Then Eq.\Ref{gh} follows once we identify $G_{AB}=G_{ab}$ and $U_{\partial f}=H_{ab}$. But Eq.\Ref{gh} characterizes the flat connection on Riemann surface. Thus the 3D discrete geometry relates to the SU(2) flat connection on Riemann surface.  

The relation between 3D discrete geometry and SU(2) flat connection on Riemann surface suggests that in the presence of cosmological constant, the phase space of LQG, consisting of the holonomies and fluxes, is equivalent to the moduli space $\cm_{flat}(\cs,SU(2))$ of SU(2) flat connections on Riemann surface.

\section{\label{esc:level4}Symplectic Structure}

The usual twisted geometry phase space $P=S^1\otimes S^1\otimes T^*S^1$ quotient out $\mathbb{Z}_2$ and the kernel of the symplectic structure is symplectomorphic to $T^*SU(2)$, which is the phase space of LQG on an edge, at the non-gauge-invariant level. But we still need to pick out the subspace fulfilling the closure condition and quotient out the $SU(2)^V$ gauge equivalence which is generated by closure condition on each tetrahedron. Finally at the gauge-invariant level, LQG phase space is $T^*SU(2)^E//SU(2)^V$. 
	

	However, in this paper we relate the SU(2) flat connection on Riemann surface to the generalized twisted geometry with curved tetrahedra. The flat connection on Riemann surface automatically take the closure condition Eq.(\ref{fun2}) into account. So the discussion in this paper is directly at the gauge invariant level. 
	
For a closed Riemann surface $\cs$ (relating to a closed graph by $\hat{\rho}$), the moduli space of SU(2) flat connections $\cm_{flat}(\cs,SU(2))$ is a symplectic space, whose symplectic structure is $\O=\frac{k}{4\pi}\int_{\cs}\tr\lt[\delta_1 A\wedge \delta_2 A\rt]$ ($k$ becomes Chern-Simons level in quantum theory). $\O$ can be derived from Chern-Simons theory on $\cs\times\R$. We propose that in the presence of cosmological constant $\L$, $\cm_{flat}(\cs,SU(2))$ is a generalization of the LQG phase space $T^*SU(2)^E//SU(2)^V$ base on the graph $\hat{\rho}(\cs)$. It is easy to check that they have the same dimension, by Eq.(\ref{trans}) translating the flat connection variables to twist geometry variables.  
	
	 What's more, the symplectic form $\O$ can be parametrized by $x_{ab}$ and $\xi_{ab}$ defined in Section \ref{G4S}, which have twisted geometry interpretations. The variables $x_{ab}$ and $\xi_{ab}$ relates to the complex Fenchel-Nielsen(FN) coordinates of flat connections (see Appendix \ref{FNcoo}). The FN coordinates are the symplectic coordinates on $\cm_{flat}(\cs,SU(2))$. As a result, the sympectic form can be expressed as
	\be
	\O=-\frac{ik}{2\pi}\rmd \xi_{ab}\wedge \rmd\ln x_{ab}+\cdots
	\ee
$\cdots$ stands for the symplectic coordinates for the flat connection on individual 4-holed spheres $S_a,S_b$. The coordinates in $\cdots$ equivalently parametrizes the shapes of constant curvature tetrahedra associated to $S_a,S_b$, which has been studied extensively in \cite{Hal.M.Haggard2015}.

Geometrically $\xi_{ab}$ and $\ln x_{ab}$ relate to the hyper-dihedral angle $\Theta_{ab}$ and area $A_{ab}$ of the interface $f_{AB}$. By using the relation $x_{ab}=\exp(\frac{2\pi i}{k}j_{ab})$, the above symplectic structure $\O$, derived from $\cm_{flat}(\cs,SU(2))$, reproduces the right Poisson bracket for the twist geometry 
\be
\{j_{ab},\xi_{ab}\}=1
\ee 
Here $\frac{2\pi}{k}\propto |\L|\ell_P^2$ and $j_{ab}\ell_P^2$ is proportional to the area $A_{ab}$ up to the ambiguity mentioned below Eq.(\ref{ambig}). 

Other canonical variables in $\cm_{flat}(\cs,\Su)$ describe the shapes of tetrahedra. They has been studied in \cite{Hal.M.Haggard2015}, and shown to be a proper generalization from the case of flat tetrahedra.

The above discuss suggests that $\cm_{flat}(\cs,SU(2))$ is indeed the right phase space of LQG or twisted geometry in the presence of cosmological constant.

\section{Quantization}\label{QUAN}

Given that $\cm_{flat}(\cs,SU(2))$ is the right phase space for LQG with cosmological constant, we would like to understand the quantization of the phase space and its implication to quantum 3d geometry. 

The quantization of $\cm_{flat}(\cs,SU(2))$ has been well-understood in the development of Chern-Simons theory with compact gauge group. $\cm_{flat}(\cs,SU(2))$ is also the phase space of Chern-Simons theory on $\cs\times\R$. See e.g. \cite{Alekseev:1994pa,Elitzur:1989nr} for the results of quantization. However instead of provide a full exposition of the quantum theory, we rather focus on quantizing the quantities which have geometrical interpretations in twisted geometry, e.g. the face area and twisted angle. 

Because the phase space $\cm_{flat}(\cs,SU(2))$ is compact, the proper coordinates of the phase space relating the area $A_{ab}$ and twisted angle $\xi_{ab}$ are the exponentials $x_{ab}$ and $y^2_{ab}=e^{-i\xi_{ab}}$ (see Appendix \ref{FNcoo} for details). The symplectic structure implies
\be
\{\ln x_{ab},\ln y_{ab}^2\}=\frac{k}{4\pi}
\ee
In quantum theory, the quantization of $x_{ab},y^2_{ab}\in U(1)$ is the same as a quantum torus. The prequantum line bundle over $U(1)\times U(1)$ has a curvature $\frac{k}{2\pi}\rmd \ln y_{ab}^2\wedge \rmd\ln x_{ab}$. Weil's integrality criterion then implies that $k\in\Z$. We choose the $x_{ab}$-polarization such that the wave function is written as $f(\ln x_{ab})$, satisfying both periodicity and invariant under Weyl reflection\footnote{The Weyl reflection $x_{ab}\to x_{ab}^{-1}$ is a redundancy of the coordinate $x_{ab}$.} $f(\ln x_{ab}) = f(-\ln x_{ab} ) = f(\ln x_{ab} + 2\pi i)$. The periodicities in both $\ln x_{ab}$
  and $\ln y_{ab}^2$ implies that $\ln x_{ab}$ can only take $k+1$ discrete values $\ln x_{ab} = 0, \frac{i\pi}{k} , \frac{2i\pi}{k}, \cdots, i\pi$, i.e.
\be
x_{ab}=e^{\frac{2\pi i}{k}j_{ab}},\ \ j_{ab}=0,\half,\cdots,\frac{k}{2}.
\ee
Given the relation between $j_{ab}$ and the area $A_{ab}$. the above implies the discrete area spectrum with a cut-off $\frac{k\ell_P^2}{2}\propto|\L|^{-1}$.

In the same way, in the $y_{ab}$-polarization where the wave function is $f(\ln y_{ab}^2)$, one find $\ln y_{ab}^2$ can only take $k+1$ discrete values $\ln y_{ab}^2 = 0, \frac{i\pi}{k} , \frac{2i\pi}{k}, \cdots, i\pi$. Given that $y^2_{ab}=e^{-i\xi_{ab}}$, we obtain a discrete spectrum of twist angle $\xi_{ab}$
\be
\xi_{ab}=0, \frac{\pi}{k} , \frac{2\pi}{k}, \cdots, \pi
\ee 
Provided the relation $\xi_{ab}=\g\Theta_{AB}$, the quantization implies a discrete spectrum of hyper-dihedral angle $\Theta_{AB}$, which is a new phenomena in the presence of cosmological constant. It might relate to the discreteness of time in LQG, as proposed in \cite{Rovelli:2015fwa}.

\section*{Acknowledgements}
The authors acknowledges Carlo Rovelli and Simone Speziale for some clarification on twisted geometry. MH acknowledges Yau Mathematical Sciences Center at Tsinghua University in Beijing for the hospitality during his visits. HM also acknowledges the support from the National Science Foundation through grant PHY-1602867, and the Start-up Grant at Florida Atlantic University, USA.

\appendix

\section{Complex Fenchel-Nielsen (FN) Coordinate}\label{FNcoo}

\begin{figure}[h]
\begin{center}
\includegraphics[width=8cm]{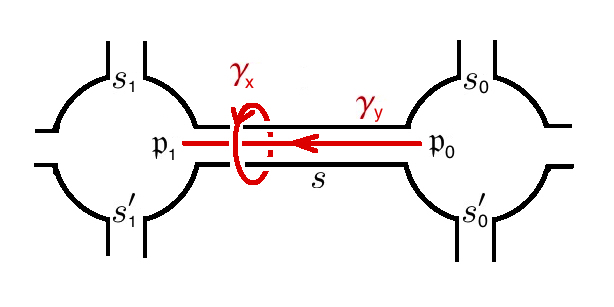}
\caption{A Riemann surface $\cs$ is stretched as two 4-holed spheres connected by a cylinder. $\g_x$ and $\g_y$ are the meridian and longitude curves of the cylinder, which is useful in defining complex FN coordinate. $s_{0,1},\ s_{0,1}'$ and $s$ denote the framing flags associate to the boundaries and cylinder. }
\label{FN}
\end{center}
\end{figure}

Consider a Riemann surface $\cs$ shown in FIG.\ref{FN}, $\cs$ may be stretched into a pair 4-holed spheres connected by a cylinder. The stretching leads to the definition of a pair of complex Fenchel-Nielsen (FN) coordinates on $\cm_{flat}(\cs,\Slc)$ the moduli space of (framed) flat $\Slc$ connections on $\cs$  \cite{kabaya,DGV}. 

A \emph{framed} $\Slc$ flat connection on $\cs$ is an $\Slc$ flat connection $A$ with a choice of flat section $s$ (called the \emph{framing flag}) in an associated flag bundle (1) over every boundary components of $\cs$ (the holes) and (2) over the cylinder connecting the two 4-holed spheres \cite{DGV,FG03}. The flat section $s$ may be viewed as a $\C^2$ vector field, defined up a complex rescaling and satisfying the flatness equation $(\rmd-A)s=0$. Each $s$ from either a boundary component or the cylinder may be extend to be a flat section on the entire $\cs$ by the flatness equation, although $s$ from different origin result in different flat sections on $\cs$. 

Let $s$ be the framing flag on the cylinder in FIG.\ref{FN}. Obviously $s$ at a point $\fp$ on the cylinder is the the eigenvector of the holonomy $H(\fp)$ along the meridian curve $\g_x$ based at $\fp$. The eigenvalue $x\in\C$ of $H(\fp)$ defines the complex FN length variable, i.e.
\be
H(\fp)=M(\fp)\left(
\begin{array}{cc}
x & 0\\ 
0 & \ \ x^{-1}
\end{array}
\right)M(\fp)^{-1}.\label{MxM}
\ee
Here $M(\fp)\in\Slc$ used for diagonalization depends on the base point $\fp$ of $H(\fp)$. The first column of $M(\fp)$ is just $s(\fp)$ up to a normalization.

Although $s$ is defined as the framing flag on the entire cylinder, from the view point of two 4-holed spheres, we are motivated to define two different framing flags by choosing two different initial values $s(\fp_0)$ and $\tilde{s}(\fp_1)$ for the flatness equation. The flat sections developed from $s(\fp_0)$ and $\tilde{s}(\fp_1)$ are denoted by $s$ and $\tilde{s}$ respectively. The two framing flags $s,\tilde{s}$ on the cylinder come from the two framing flags associated to the two holes of 3-holed spheres connected by the cylinder. We find that $s$ and $\tilde{s}$ are only different by a rescaling, which is denoted by $y\in\C$. Indeed, we denote by $G(\fp_1,\fp_0)$ the holonomy along $\g_y$ traveling from $\fp_0$ to $\fp_1$. Then\footnote{It is easy to observe that $G(\fp_1,\fp_0) H(\fp_0)=H(\fp_1) G(\fp_1,\fp_0)$ by the flatness of the connection $A$. By Eq.\Ref{MxM}, we have
\be 
&&M(\fp_1)^{-1}G(\fp_1,\fp_0) M(\fp_0)\left(
\begin{array}{cc}
x & 0\\ 
0 & \ \ x^{-1}
\end{array}
\right)\nonumber\\
&=&\left(
\begin{array}{cc}
x & 0\\ 
0 & \ \ x^{-1}
\end{array}
\right)M(\fp_1)^{-1}G(\fp_1,\fp_0)M(\fp_0).
\ee
where the first columns of $M(\fp_0),M(\fp_1)$ are $s(\fp_0),\tilde{s}(\fp_1)$. It implies the diagonalization $M(\fp_1)^{-1}G(\fp_1,\fp_0)M(\fp_0)=\mathrm{diag}(y,y^{-1})$. Eq.\Ref{Gss} is obtained by restricting the attention to the first column of $M(\fp)$'s.}
\be
G(\fp_1,\fp_0) s(\fp_0)=y \tilde{s}(\fp_1).\label{Gss}
\ee
It is clear that the parameter $y$ depends on the choices and normalizations of $s(\fp_0)$ and $\tilde{s}(\fp_1)$.

As the variable canonical conjugate to $x$, the complex FN twist $\t$ is defined in the following way: Consider a longitude curve $\g_y$ traveling along the cylinder connecting two points $\fp_0,\fp_1$ on two different 4-holed spheres (see FIG.\ref{FN}). Let $s_{0,1},s_{0,1}'$ be the framing flags for 2 pairs of boundary components in $\partial \cs$, $\t$ is defined by (see e.g. \cite{DGV})\footnote{For convenience, $\t$ defined here is different to the one in \cite{DGV} by a minus sign}
\be
\t=\frac{( s_{0}\wedge s_{0}')}{( s_{0}\wedge s)( s_{0}'\wedge s )}\frac{( s_{1}\wedge s)(s_{1}'\wedge s)}{( s_{1}\wedge s_{1}')}.\label{tau}
\ee 
where the $\Slc$ invariants $s\wedge s'$ are evaluated at a common point after parallel transporting $s$ and $s'$. Without loss of generality, we evaluate the first ratio with factors ${( s_{0}\wedge s_{0}')},( s_{0}\wedge s),( s_{0}'\wedge s )$ at $\fp_0$, and evaluate the second ratio with factors ${( s_{1}\wedge s),(s_{1}'\wedge s)},{( s_{1}\wedge s_{1}')}$ at $\fp_1$. The evaluation involves both $s(\fp_0)$ and $s(\fp_1)$ at two ends of $\g_y$, while the parallel transportation between $s(\fp_0)$ and $s(\fp_1)$ depends on a choice of contour $\g_y$ connecting $\fp_0,\fp_1$. Different $\g_y$ may transform $s(\fp_1)\to x s(\fp_1)$. 

One can show that $x,\t$ are canonical conjugate variables of the holomorphic Atiyah-Bott-Goldman symplectic form $\O=\int_{\cs}\tr\lt[\delta_1 A\wedge \delta_2 A\rt]$, i.e. the reduction of the symplectic from to $x,\t$ gives \cite{GMN09,2CCS}
\be
\O=\frac{\rmd \t}{\t}\wedge\frac{\rmd x}{x}+\cdots.
\ee

We evaluate the twist variable $\t$ by evaluating the first ratio in Eq.\Ref{tau} at $\fp_0$, while evaluating the second ratio at $\fp_1$:
\be
\t&=&\frac{( s_{0}(\fp_0)\wedge s_{0}'(\fp_0))}{( s_{0}(\fp_0)\wedge s(\fp_0))( s_{0}'(\fp_0)\wedge s(\fp_0) )}\nonumber\\
&&\frac{( s_1(\fp_1)\wedge G(\fp_1,\fp_0) s(\fp_0))(s_{1}'(\fp_1)\wedge G(\fp_1,\fp_0) s(\fp_0))}{( s_{1}(\fp_1)\wedge s_{1}'(\fp_1))}\nonumber\\
&=&y^2\Big[\frac{( s_{0}(\fp_0)\wedge s_{0}'(\fp_0))}{( s_{0}(\fp_0)\wedge s(\fp_0))( s_{0}'(\fp_0)\wedge s(\fp_0) )}\nonumber\\
&&\frac{( {s}_1(\fp_1)\wedge \tilde{s}(\fp_1))(s_{1}'(\fp_1)\wedge \tilde{s}(\fp_1))}{( s_{1}(\fp_1)\wedge s_{1}'(\fp_1))}\Big].\label{tauy}
\ee
The quantity in the bracket only depends on the flat connections and framing flags located in the pair of 4-holed spheres. The data of framing flags are assumed a priori. Only the holonomy $ G(\fp_1,\fp_0)$ traveling from one 4-holed sphere to the other has a nontrivial intersection with meridian holonomy $H$, while $H$ essentially has no intersection with any holonomy located in a single 4-holed sphere. Therefore the quantity in the bracket Poisson commutes with $x$, so we can rewrite the symplectic structure as
\be
\O=\frac{\rmd y^2}{y^2}\wedge\frac{\rmd x}{x}+\cdots.
\ee 

We introduce on $\C^2$ the Hermitian inner product $\lag s,s'\rag=\bar{s}^1{s'}^1+\bar{s}^2{s'}^2$, and normalize $s(\fp_0)$ and $\tilde{s}(\fp_1)$ by $\lag s(\fp_0),s(\fp_0)\rag=\lag \tilde{s}(\fp_1),\tilde{s}(\fp_1)\rag=1$. Under this normalization, we restrict our attention to SU(2) flat connections as a subspace in $\cm_{flat}(\cs,\Slc)$, and we want to understand what is the restriction of the variables $x,y$. Firstly, it is obvious that $x\in \mathrm{U}(1)$. The matrices $M(\fp_0),M(\fp_1)\in\mathrm{SU(2)}$ are written as
\be
M(\fp_0)&=&\left(
\begin{array}{cc}
s^1(\fp_0) & -\bar{s}^2(\fp_0)\\ 
s^2(\fp_0) & \ \bar{s}^1(\fp_0)
\end{array}\rt),\nonumber\\
M(\fp_1)&=&\left(
\begin{array}{cc}
\tilde{s}^1(\fp_1) & -\bar{\tilde{s}}^2(\fp_1)\\ 
\tilde{s}^2(\fp_1) & \ \bar{\tilde{s}}^1(\fp_1)
\end{array}\rt)
\ee
The flat connection being SU(2) implies $G(\fp_1,\fp_0)\in \mathrm{SU(2)}$, then implies $y\in \mathrm{U(1)}$. We have written $y= e^{-i\xi/2}$ in Eq.(\ref{dia2}).


\end{document}